\documentclass[letterpaper,twocolumn,10pt]{article}
\usepackage{usenix2019_v3.1}

\usepackage{tikz}
\usepackage{amsmath}

\usepackage{filecontents}

\usepackage{textgreek}
\usepackage{hyperref}
\usepackage{graphicx}
\usepackage{comment}
\usepackage{authblk}

\title{\LARGE \bf
Cognitive Computing in Data-centric Paradigm
}

\author[1]{Viacheslav Dubeyko}

\begin{document}

\maketitle
\pagestyle{plain}

\begin{abstract}

Knowledge is the most precious asset of humankind. People extract the experience from the data that provide for us the reality through the feelings. Generally speaking, it is possible to see the analogy of knowledge elaboration between humankind's way and the artificial system's way. Digital data are the "feelings" of an artificial system, and it needs to invent a method of extraction of knowledge from the Universe of data.

The cognitive computing paradigm implies that a system should be able to extract the knowledge from raw data without any human-made algorithm. The first step of the paradigm is analysis of raw data streams through the discovery of repeatable patterns of data. The knowledge of relationships among the patterns provides a way to see the structures and to generalize the concepts with the goal to synthesize new statements. The cognitive computing paradigm is capable of mimicking the human's ability to generalize the notions. It is possible to say that the generalization step provides the basis for discovering the abstract notions, revealing the abstract relations of patterns and general rules of structure synthesis.

If anyone continues the process of structure generalization, then it is possible to build the multi-level hierarchy of abstract notions. Moreover, discovering the generalized classes of notions is the first step towards a paradigm of artificial analytical thinking. The most critical possible responsibility of cognitive computing could be the classification of data and recognition of input data stream's states. The synthesis of new statements creates the foundation for the foreseeing the possible data states and elaboration of knowledge about new data classes by employing synthesis and checking the hypothesis.

\end{abstract}



\section{INTRODUCTION}

Cognitive computing still looks like a fantasy yet, and anyone could claim to provide any evidence of the advantage of the new paradigm comparing with the old one. Why and where could anybody use cognitive computing? How feasible and applicable this computing paradigm for data processing? Nowadays, it's hard to say that cognitive computing is a bright idea, and anyone could misunderstand the background, the goal, and possible outcomes. However, it is possible to believe that cognitive computing has a fantastic future and is capable of breaking through the data processing performance. At first, it needs to figure out the possible directions to employ cognitive computing and potential issues of using it.

\section{WHY COGNITIVE COMPUTING?}

\subsection{Big Data problem}

The Big Data problem inspires every Computer Scientist during the last several decades. Moreover, every year this problem becomes more crucial under exponential growth of available data capacity and more profound diversity of unstructured data. The modern data processing paradigm has achieved the limits of technological potential and is unable to provide the next level of breakthrough in data processing performance. It means that advanced computing technologies are unable to execute analytical requests in the environment of modern time restrictions because of the wide variance of data types and a massive amount of available data. Distributed nature of open data, enormous variety of data types, continuous and fast evolution of users' requests are the basis of crucial need in elaboration of a new computing paradigm can evolve quickly, be ready to operate in genuinely decentralized environment, be capable to estimate the relevancy of data under analysis, be able to "synthesize" some conclusions. The capability to synthesize the findings could forecast or calculate the probability of a possible future state of the information stream. Also, it can be used for generalization of notions to classify data in saved information streams. Cognitive computing is the paradigm that could achieve a breakthrough in the performance of Big Data processing.

\subsection{CPU-centric paradigm inefficiency and memory wall problem}

Nowadays, the CPU-centric paradigm is a crucial bottleneck for the fundamental improvement of data processing performance. From one point of view, there is a naive direction of performance improvement through increasing the number of CPU's cores. But it is not a feasible solution because of the significant amount of bottlenecks and drawbacks that creates this approach. CPU-centric paradigm was born by scientists to automate the execution of algorithm-based tasks. To be more precise, the goal was to transform a mathematical formula into an algorithm that can be executed automatically many times without involving the people. Generally speaking, as a result, the fundamental basis of such an approach plays a code of operation that defines as essence of activity as location and type of data under supervision. It needs to point out that the CPU-centric approach was a fundamental breakthrough at that time. It created the whole eco-system of automatic data processing based on algorithms. Moreover, a high-level programming language is building a tool for managing the complexity of the problem.

Moreover, the available volume of existing digital data was so negligible, and it was tough to imagine how data itself can define the execution of data transformation at the time of the CPU-centric paradigm invention. As a result, any performance in the environment of the CPU-centric paradigm is impossible without delivering some implemented algorithm. The fundamental drawback of this paradigm is the inevitable necessity to transfer as a code of operation as data itself into an execution core. Moreover, CPU's registers stack was the brilliant vision at the dawn of computing paradigm creation, but nowadays, it converted into the bottleneck that is called a memory wall problem. It means that operation can be executed only inside the execution core, and a limited set of registers look like the eye of a needle that connects the execution core and main memory. Generally speaking, it is possible to expect a breakthrough in data processing performance only by exchanging the CPU-centric paradigm by decentralized data processing inside the persistent data storage. One of the promising directions could be the cognitive computing paradigm that is capable of revolutionizing the data processing performance.

\subsection{Unstructured data problem}

Any algorithm needs input data for execution. Also, usually, it generates some digital data as a result of performance. The invention of computing technologies had introduced the problem of storing and ordering of digital data. As a result, computer scientists invented many data formats of representation and storing digital data. However, digital data needs to be not only saved but also to be accessed. Generally speaking, it implies to have such relations between data that provide the opportunity to execute some requests to extract relevant data. Relational databases created a potent tool for the execution of such claims. However, the relational data model needs in designing a database's scheme, a tool for adding and cleaning the data, and the development of complex SQL requests. All these requirements create critical obstacles for more wide using the relational data model in real life and, finally, decrease the worth and efficiency of this data model. The problem of a proper data model is getting more and more crucial in the modern world by increasing the diversity and volume of available data. This challenge has created a very dynamic area of NoSQL databases and unstructured data model of storing and manipulation by data. Unstructured data problem needs in a flexible model of data organization, and order is quickly evolving with new data type introduction. Generally speaking, the cognitive computing paradigm can be a very flexible approach that can manage the unstructured data problem.

\subsection{Massively parallel data processing problem}

Scientists recognized from the dawn of Computer Science that parallel calculations could be a compelling resource for the significant improvement of the data processing. However, CPU-oriented data processing creates a lot of problems and obstacles for massively parallel data processing. First of all, it needs to point out that it is not so easy to split efficiently any algorithm into several parallel threads for faster processing of shared data. As a result, the CPU-centric model is fundamentally not efficient for the case of massively parallel data processing. Another severe problem of this approach is the necessity to synchronize access of several threads to shared data and to implement the complex coherence protocols to synchronize the state of data between CPU's caches' and main memory (DRAM). Generally speaking, the CPU-centric approach has a significant number of drawbacks for the case of massively parallel data processing, and it is hard to consider it as a proper computing technology for the next technological step of humankind. However, a cognitive computing paradigm is capable of using a fundamentally orthogonal scheme of data processing than the CPU-centric model is using. For the case of cognitive computing, data itself can be a fundamental basis that initiates, construct, and program the calculation or execution flow. It means that the cognitive computing paradigm does not need to use an algorithm designed by a human for data processing. This computing paradigm is responsible for elaborating a data processing logic is relevant to the current state of an input data stream. Generally speaking, decentralized data processing and analysis could employ a massively parallel model of data processing that starts from an independent investigation of data portions by different processing cores and finishes by notions generalization. These generalized notions could build the fundamental basis for further analysis of input data stream, constructing and checking hypothesis. Finally, processing cores can communicate, collaborate, and make more complex structures to analyze and process the data.

\subsection{Algorithm-oriented data processing bottlenecks}

Nowadays, the modern computing paradigm requires an algorithm defined in the form of executable instructions before any data transformation. The code of operation establishes the type of data, location of data, and type of data transformation in the Turing machine paradigm. Generally speaking, this fundamental requirement creates critical side effects that cannot be resolved by simple polishing and improvement of the initial idea. First of all, any algorithm includes the suppositions related to a type of processing data. It means that a new kind of data needs in the development of a new specialized algorithm.
Moreover, very frequently, an algorithm takes into account the peculiarities of a hardware platform that results in the inability to port the algorithm easily into other hardware platforms. Also, usually, one algorithm's step depends on the results of the previous one in the whole sequence. Generally speaking, the sequential nature of an algorithm's execution is a complicated environment for parallel running the different steps or iterations of an algorithm. The same critical issue is the necessity to share data between iterations or threads that implement an algorithm. It needs to employ the synchronization primitives (semaphores, mutexes, and so on) for managing the access to shared data, and it could result in performance degradation or potential race conditions and deadlocks. Finally, any algorithm-oriented calculation process is sequential applying of operations sequence for a set of data of the same type. Nowadays, modern computing technologies originate more and more new data types that need to process in a faster and reliable way. Generally speaking, the algorithm-oriented paradigm is unable to manage the modern data processing requirements by the lack of enough qualified human staff and complexity to develop, maintain, and test the complex program systems. Complex program systems are hard to create, to test, and to maintain. Such systems are not flexible and are unable to evolve or to change architecture easily. Finally, the cognitive computing paradigm could play the role of an alternative computing paradigm that can grow based on available data. This paradigm is capable of being very flexible and evolving in the environment of continuously changing unstructured data that reflect the state of evolving reality.

\subsection{AI-oriented data analysis and cognition problem}

The available data volume is growing exponentially every day. Moreover, a variety of data types is increasing, which makes the problem of data analysis more crucial than ever before. Such a critical environment requires to use of AI-oriented methods of data analysis. It means that the reality of data processing requires to employ more flexible approaches can evolve for continuously changing everyday tasks. Finally, a cognition problem needs to consider the cornerstone problem. It needs to point out that the challenge of universal AI is far from to be solved soon. This problem contains many not addressed yet philosophical issues. However, the current state of the art is ready to provide the technologies that could be a basis for the creation of an artificial system with cognition features. Such a cognition feature can be the foundation for recognition of data nature and self-elaboration of data processing strategy that could serve to the needs of a user or another artificial system. Generally speaking, the naive view on cognition feature is the ability of an artificial system to recognize the repeatable patterns and structures of it in the raw data stream. Finally, such capability of patterns and structures recognition converts the problem of data analysis into a decentralized method that will be able to generalize the notions and to classify the raw data of input stream based on knowledge of generalized concepts. As a result, the classification of input data could be the basis for the elaboration of the system's strategy of behavior in the current environment. Moreover, the cognitive computing paradigm could be the fundamental foundation that is capable of implementing AI-oriented data analysis.

\subsection{Continuous data streams, analytical conclusion, and deduction problems}

All real processes have a continuous and ever-changing nature. It means that such a method generates a constant information stream. Finally, the information system has to analyze data and to make conclusions under the continuous pressure of newly received data. This ever-changing nature of live data makes the classical CPU-oriented paradigm inefficient and very complicated for such data analysis. Generally speaking, this class of data analysis needs in completely different computing paradigm. The cognitive computing paradigm can solve the problem of study the ever-changing data. First of all, this paradigm provides a way of massively parallel computing via a decentralized model of data analysis. Also, secondly, cognitive computing can analyze data without any algorithm using the cognition feature. Moreover, it is possible to synthesize and to check the hypothesis in the environment of cognitive computing that provides the basis of a more profound generalization of notions and the new knowledge elaboration.

\section{WHAT IS COGNITIVE COMPUTING?}

\subsection{Digital data universe like a challenge}

Knowledge is the most precious asset of humankind. People extract the experience from the data that provide for us the reality through the feelings. Generally speaking, it is possible to see the analogy of knowledge elaboration between humankind's way and the artificial system's way. Digital data are the "feelings" of an artificial system, and it needs to invent a method of extraction of knowledge from this Universe of data.

During the dawn of the creation of computing technologies, the user data volume was comparable or lesser than the amount of program code. It was more important to suggest the automated implementations of algorithms at those time. We are living in an era of algorithm-oriented computing because of this reason. However, scientists realized the importance of user data while the first databases were inventing. Moreover, the case of distributed databases crystallized the crucial point and deep complexity of data management.

Generally speaking, the real boom of data generation has taken place with the creation of personal computers that make the digital data by the reality of everyday life of ordinary people. Cloud storage technologies created another great revolution. These technologies provide the opportunity to store and to access the enormous volume of digital data practically from any place in the world. Finally, the cloud storage technologies have made the Big Data by the reality of everyone. Moreover, the Big Data have revealed the crucial inefficiency of modern computing paradigm. Generally speaking, the existing computing paradigm is unable to process and to analyze the available data in the environment of the CPU-centric computing. The Big Data problem requires a fundamentally different computing paradigm that can offer a more efficient and faster way of data processing.

\subsection{Current state of the art of hardware and software}

Nowadays, the algorithm-oriented model of the Turing machine is the computing paradigm of the modern hardware stack. It means that a raw binary stream represents any data is capable of being analyzed by an algorithm. Any algorithm is a sequence of commands that can be recognized and executed by processing cores. Generally speaking, a code of operation is the essential item that identifies an essence of action, data type, and data location. Any processing core is a set of registers and circuitry that can recognize and execute some set of commands. First of all, anyone needs to store code of operation and the related data into registers. Secondly, it will be possible to initiate the execution of an operation by circuitry under data in the registers. At last, the chip stores the result of activity into the designated register that flushes into the main memory afterward.

Fast computing core needs in fast memory that, usually, is built as SRAM cache in modern CPUs. But SRAM is expensive, and the capacity of L1/L2/L3 cache is small. As a result, the slower main byte-addressable and volatile memory is keeping the executing code and data under operation. However, the main memory is volatile, and it cannot store data persistently. Finally, slow persistent and block-addressable memory plays the role of long-term storage of code/data that keeps the data between system restarts. Generally speaking, fundamentally, the CPU-centric computing paradigm implies the enormous amount of moving operations between the processing cores' registers and the main memory.

Moreover, all types of memory do not know data nature and are not capable of fulfilling any transformation with the raw binary stream of data. The magical point takes place in the processing core, while code is associating with data. Generally speaking, this association creates the knowledge of data and operation's type for the duration of the operation execution.

Also, the unity of hardware and software stacks is the essence of the implementation of the computing paradigm of automated calculations.  From one viewpoint, the advantage of this solution is the opportunity to evolve by hardware and software stacks independently. Generally speaking, it means that different hardware platforms can use the same software stack in the presence of hardware-specialized drivers. Oppositely, the software stack can be modified if it keeps the interface and protocol of interaction with hardware unchanged. The software stack is the unity of kernel space (OS activity) and user-space (application activity). Generally speaking, the responsibility of OS is to represent a model of an abstract machine that can be implemented by various hardware platforms and to manage the available resources. Oppositely, the responsibility of the application layer is to interact with a user and to execute the tasks of data management and data processing. Fundamentally, a thread is a vital concept that is responsible for data processing utilizing memory allocation and using the time slices of shared CPU core. Generally speaking, threads live in the main memory, and they represent temporary containers that keep a copy of data from persistent memory while CPU core is processing data in the allocated memory. As a result, thread's context has to be loaded by the CPU core whenever the thread receives the time slice for execution. Additionally, the CPU core has to load the code and data from the main memory. Because, otherwise, it is impossible to execute an algorithm that is implemented by a thread under execution. As a result, the massive amount of moving operations is the inevitable side effect of the CPU-centric computing paradigm.

Generally speaking, a modern computing paradigm evolved in a crucial bottleneck that cannot be resolved by merely polishing the current architecture by complete exhaustion of the paradigm's potential. Nowadays, it needs to invent a new computing and data processing paradigm that should be data-oriented and efficient for the Big Data case.

\subsection{Cognitive computing paradigm like a new way of data analysis}

It is possible to state that the most crucial issue of the modern computing paradigm is the necessity to move the data and the code from the main memory into the shared CPU core and the results of data processing in the opposite direction (into the main memory). Generally speaking, this activity results in a massive amount of moving operations (about 80\%) and a significant amount of heat generation in modern computing systems. The enormous amount of available data and exponential growth of a new data generation in persistent data storage has created a new challenge of data processing. The modern computing paradigm has exhausted the potential for improvement in the performance of data processing. It is possible to state that the lack of computing power, inability to process the data in a massively parallel way, and the necessity to spend the enormous amount of energy for data movement are the reasons for the fundamental inefficiency of the modern computing paradigm.

Generally speaking, the current computing model desperately needs to deliver code/data to the computing core that can transform data. And this way of data processing is unable to solve the problems of modern life. It needs to change the computing paradigm fundamentally. One of the possible ways is to offload data processing into the persistent storage space. However, the offload approach originates more questions than answers. First of all, the modern computing paradigm requires to deliver the binary code of algorithm into the computing core because the code of operation defines the operation's essence and the data type. Generally speaking, it implies that the efforts to save the Turing machine inevitably result in delivering the binary code into the persistent storage space even if the data transformation takes place inside a persistent memory.

Moreover, it makes sense to point out that any persistent memory consists of ample address space, and it will need to deliver the executable code, for example, to every memory page in the case of data processing offload. Finally, it sounds like a fundamentally unsolvable and unmanageable problem.  As a result, it is possible to assume that a new computing paradigm has to exclude the necessity to deliver code into the persistent space where data processing will take place. The data itself should synthesize the particular processing algorithm based on data nature and input stream's content. Generally speaking, it needs to add a cognition function into the foundation of data processing. Such cognition function would be responsible for the recognition of data nature and elaboration on a strategy of data analysis and transformation. Finally, the cognition function would make it possible to distribute the data processing in the whole persistent memory's space without the necessity to deliver a particular code to the place of data processing.

Moreover, the operation of storing or modification a data can trigger the cognition function. As a result, cognition function will work as a finite state machine that is capable of analyzing and synthesizing new data. The critical point of cognition function is the capability to distribute the computing cores through the whole persistent memory's space and to achieve the deepest possible way of massively parallel data processing and analysis. Moreover, it will be the responsibility of every computing core to "recognize" the nature and structure of available data without the necessity to deliver an algorithm for data processing. The cognition process could elaborate on the strategy of possible data analysis and to synthesize new data that would represent the analytical conclusions. Finally, the responsibility of computing cores' matrix in persistent memory space is interaction and collaboration to generalize the particular specimen of data patterns into the generalized notions.

Generally speaking, the computing cores in the persistent memory realize the cognition function is the foundation of the cognitive computing paradigm. The main activity of such computing cores is the recognition of repeatable patterns and structures in available data. As a result, the recognized patterns and structures are the basis for the generalization of notions through exchanging by particular items in the nodes of structures on a generalized concept. This abstract notion can hold any specific pattern detected in the environment. It is possible to say that the critical responsibility of the cognition function is to generalize the data to abstract notions that can classify the available data.

Moreover, a very crucial point of the cognition function is the absence of an algorithm that is developed by humans. The computing cores implement the cognition function on the fly in a decentralized manner without any necessity in centralized management. As a result, the responsibility of every computing core is an independent and decentralized elaboration of generalized notions. These elaborated concepts build the hierarchy of ideas with the concrete as leaf nodes of the tree. Finally, the hierarchy of generalized notions is used by computing cores for developing the strategy of analysis of input data and the behavior of the system at the whole as the reaction on reality evolution.

\subsection{Vision of system is implementing the cognitive computing paradigm}

The cognitive computing paradigm implies that a system should be able to extract the knowledge from raw data without any human-made algorithm. First of all, it means that the foundation of architecture has to be the persistent layer is keeping raw data. The system should be able to analyze the unstructured data, and, finally, it implies that the most generic representation of data could be the binary stream. Generally speaking, it means that data can be stored as a binary stream that should be structured by the system based on internal data nature.

\begin{figure}[h]
\centering
 
 \includegraphics[width=0.80\columnwidth,keepaspectratio]{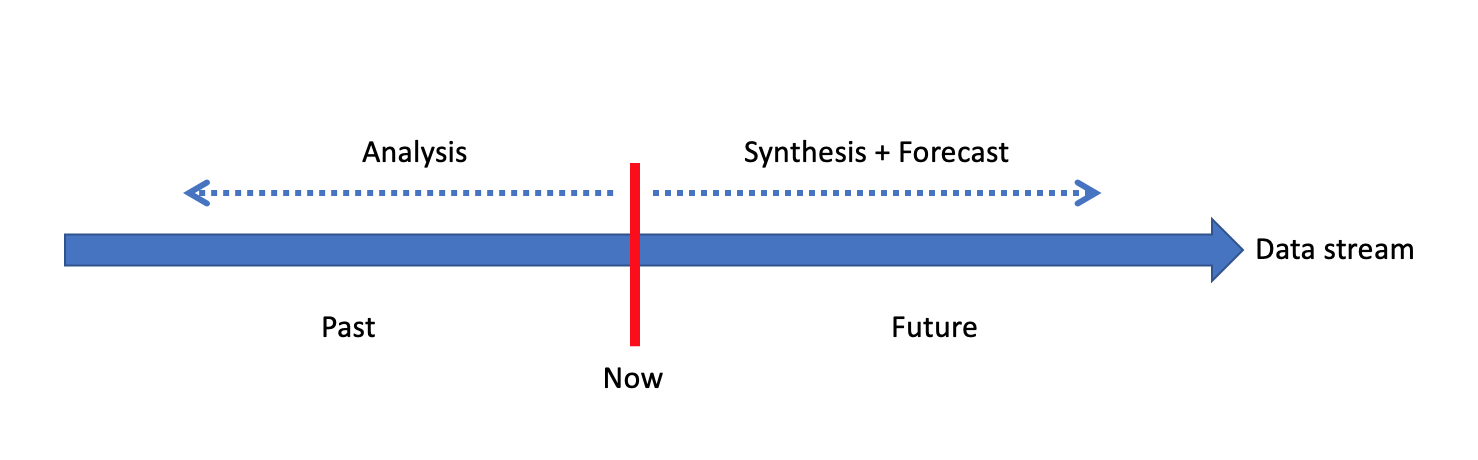}
\caption{Data stream representation.}
\label{fig:fig-00001}
\end{figure}

\begin{figure}[h]
\centering
 
 \includegraphics[width=0.80\columnwidth,keepaspectratio]{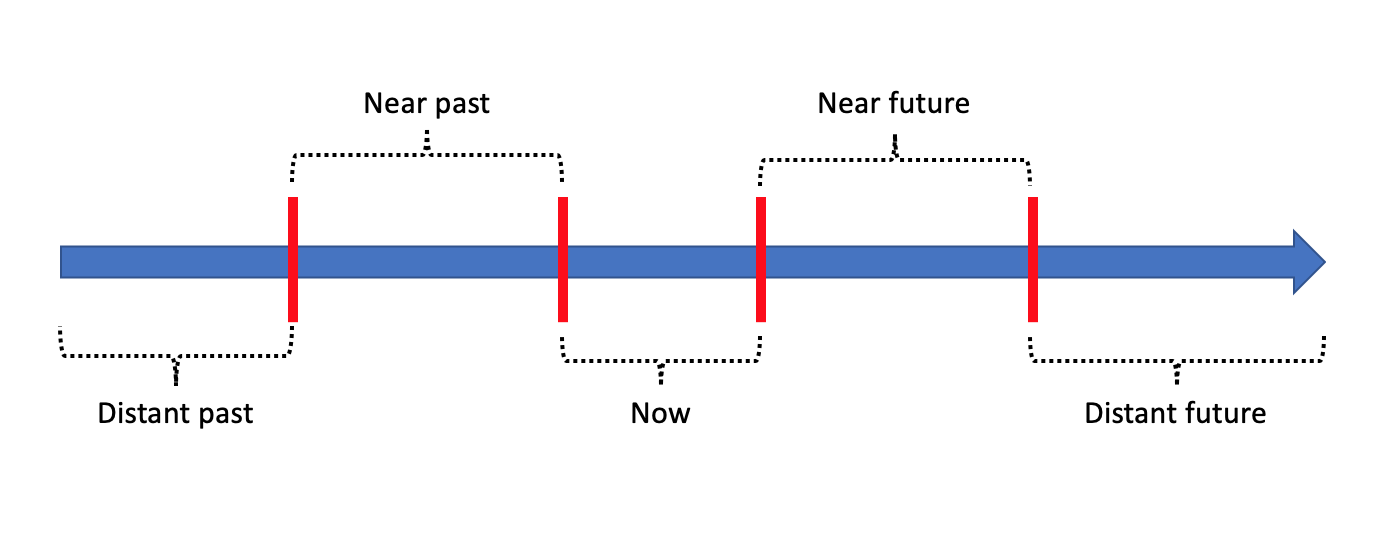}
\caption{Data stream states.}
\label{fig:fig-00002}
\end{figure}

The next level of the system has to be the cognition layer that is responsible for the recognition of repeatable patterns in the raw data by applying the cognition function. It is possible to state that the critical goal of the cognition layer is the extraction of keywords and notions in fully unstructured data under analysis. Generally speaking, it means that the cognition layer analyses a raw stream of bytes and recognizes the keywords. Also, it is possible to realize the relation of found keywords employing structures that can be represented by a graph, for example. However, the system's recognition power could depend on data nature and existing knowledge base of keywords and generalized notions. As a result, the pattern recognition subsystem can detect some patterns without "understanding" the relations amongst the patterns and essence of recognized patterns. Generally speaking, the recognition subsystem's output is some subset of patterns distinguished in the raw binary stream.

\begin{figure}[h]
\centering
 
 \includegraphics[width=0.80\columnwidth,keepaspectratio]{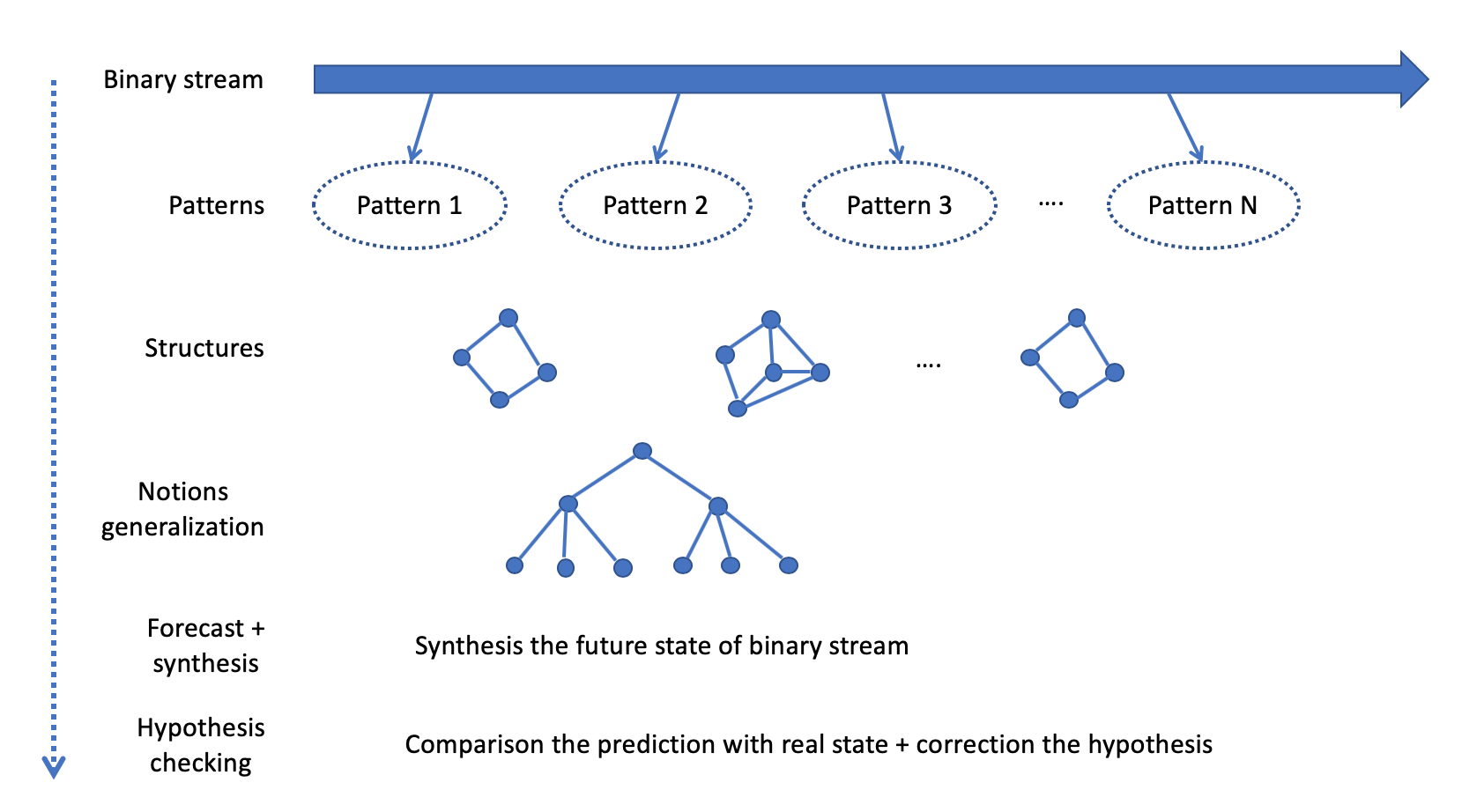}
\caption{Cognitive computing workflow.}
\label{fig:fig-00003}
\end{figure}

The next step of the analysis of raw data has to be the recognition of patterns' relations in the available raw binary streams. This task can be the responsibility of a generalization layer that should detect the similar structures and generalize particular keywords to abstract notions. In the naive approach, the notion generalization looks like changing the specific pattern on an abstract node declaration that can be substituted by any keyword from a vector or a set of keywords/patterns. Finally, several positions in the same structure could be the abstract nodes that provide a way to build the hierarchy of abstract notions. Moreover, the keyword could be converting to more generic abstraction by traversing to the root of the hierarchical tree. Generally speaking, the leaf items of such a hierarchical tree are the part of structures extracted from the raw binary stream during the recognition. Structures with one abstract keyword build the first level of abstraction in the generalization tree. As a result, the top layer of the hierarchy creates completely generic structures that contain only generalized notions.

\begin{figure}[h]
\centering
 
 \includegraphics[width=0.80\columnwidth,keepaspectratio]{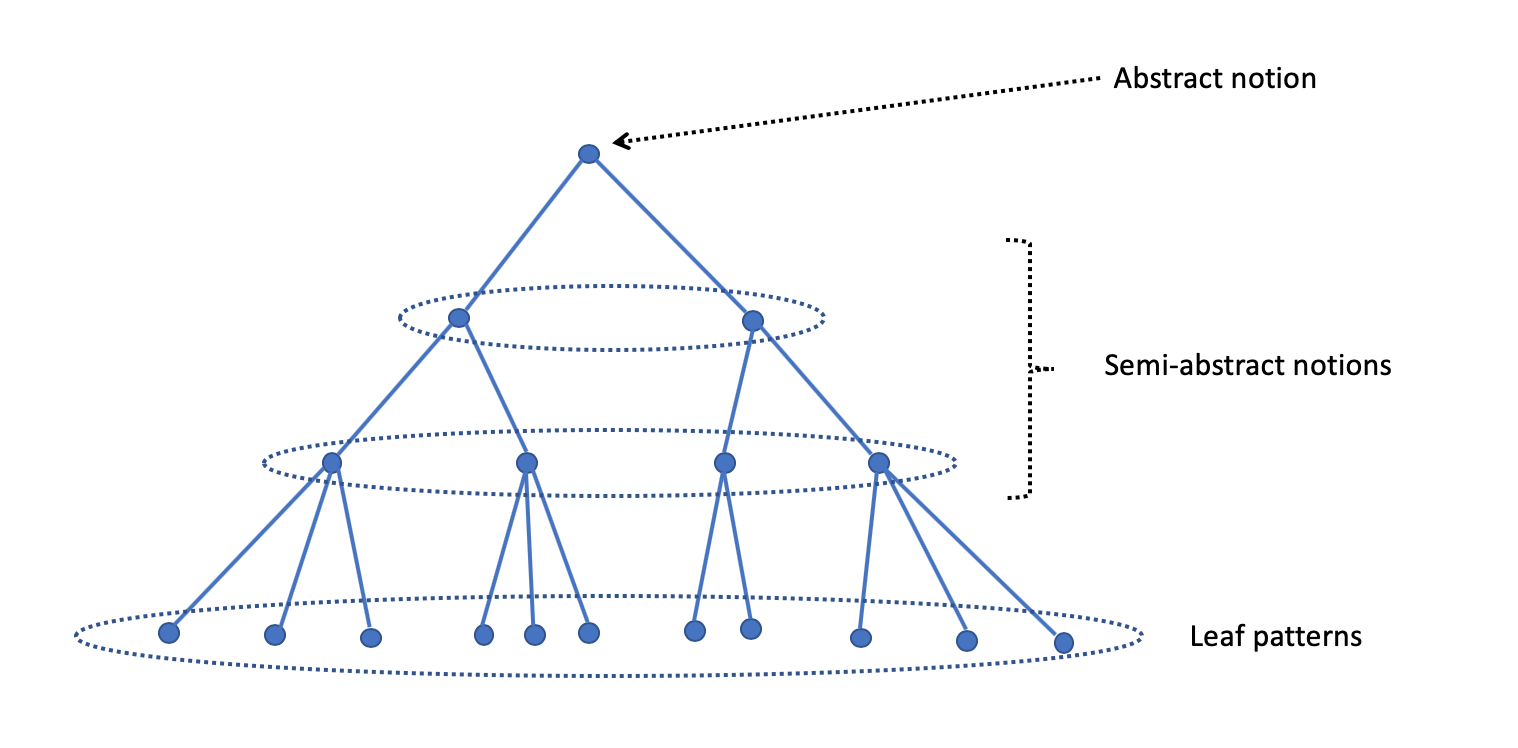}
\caption{Generalization representation.}
\label{fig:fig-00004}
\end{figure}

It is possible to see the similarity between the suggested notions hierarchy and the relational model of data. However, the proposed idea does not inherit the drawbacks and strict limitations of the relational model. First of all, the system creates the hierarchy of notions, while the raw binary stream is recognizing. Secondly, nobody needs to create any scheme before the recognition process because the system creates the hierarchical tree on the fly. And the structure of the notions hierarchy will be defined by data in the raw binary stream. The critical similarity between the suggested approach and relational model takes place because every abstract node of hierarchy gathers the knowledge about the similar structures that are distinguished by the variance of keywords in some position. Finally, it is possible to state that the tree of abstract notions mimics the human way of generalization of particular senses to generalized and abstract ideas. As a result, the generalization layer creates the mechanism of generalization and elaboration of the notions by machine itself through the recognition of repeatable patterns and structures in the raw binary stream and ordering the recognized structures into the hierarchy of abstract ideas.

Suggested subsystems are capable of elaborating some generalized notions for data ordering and execution of the SQL-like requests with the goal of data search and data extraction. Moreover, a system will be skilled enough to distinguish the repeatable patterns and to use this knowledge for parsing and analyzing the input streams with raw data. However, it will not be enough for the synthesis of new knowledge. Generally speaking, the problem of knowledge synthesis needs in the introduction of new additional layers. From one point of view, the input stream of data represents a state of reality that is continuously evolving and modifying its state. It means that the system is capable of trying to forecast the future states of input stream based on recognized patterns and generalized structures in the previously processed binary data. Another foundation of knowledge synthesis could be the users' requests that usually look like a set of keywords. Such a set of keywords can define the relevant direction of the users' interests for the new knowledge synthesis. Generally speaking, the new knowledge synthesis could look like as placing into a generalized position of structure some new patterns that are not in the known vector for this position. But these new patterns could be in some relation with the vector of recognized and registered notions. Finally, it is possible to suggest some synthesis layer that is capable of generating hypotheses in the form of synthesized structures. These structures look like a forecast of the future states of the input stream. The synthesis layer needs to be supplemented by a subsystem of hypotheses checking that could be responsible for searching the similarity of hypothesis with the states of the input stream.

\begin{figure}[h]
\centering
 
 \includegraphics[width=0.80\columnwidth,keepaspectratio]{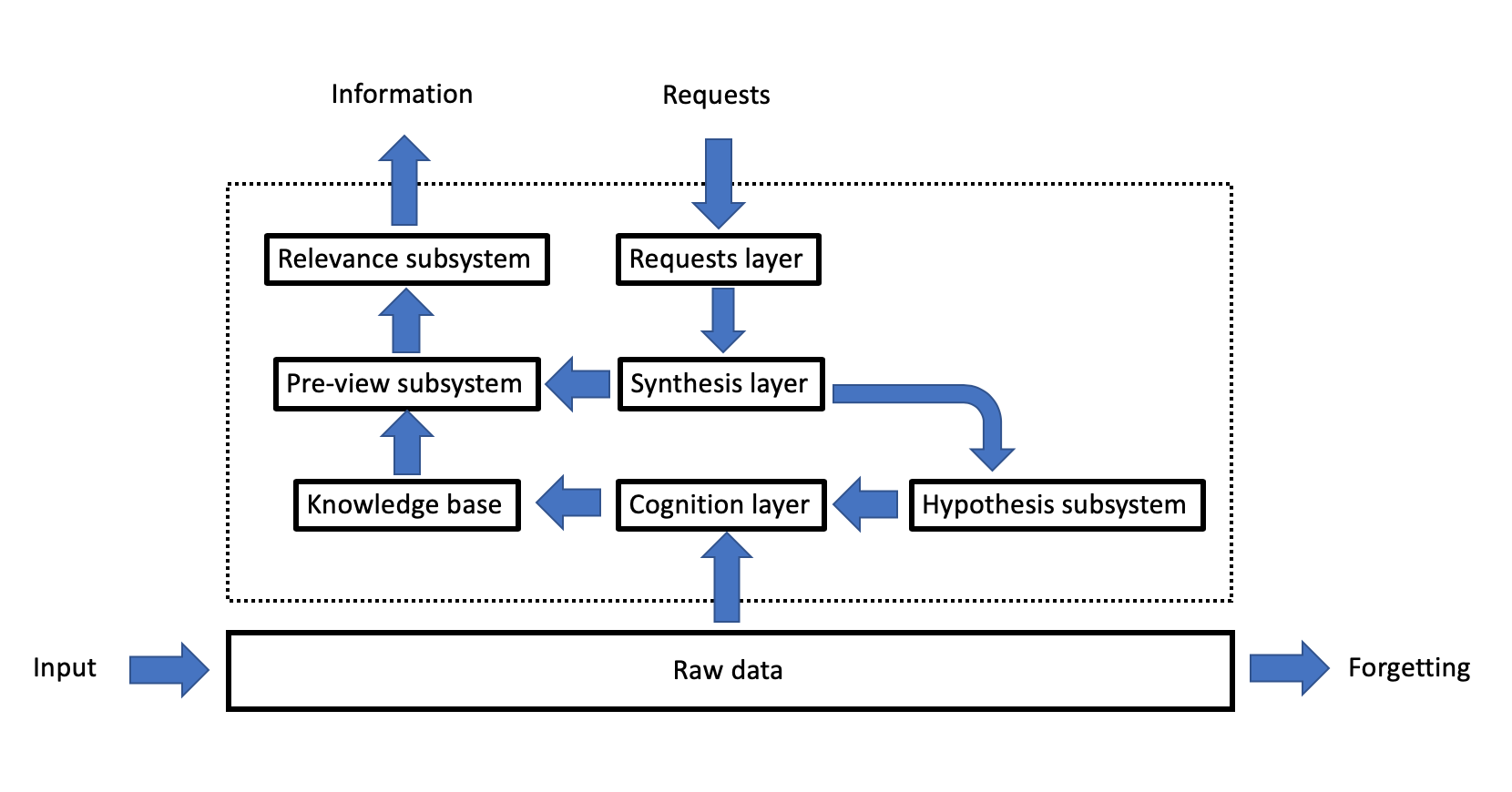}
\caption{Cognition function vision.}
\label{fig:fig-00005}
\end{figure}

\begin{figure}[h]
\centering
 
 \includegraphics[width=0.80\columnwidth,keepaspectratio]{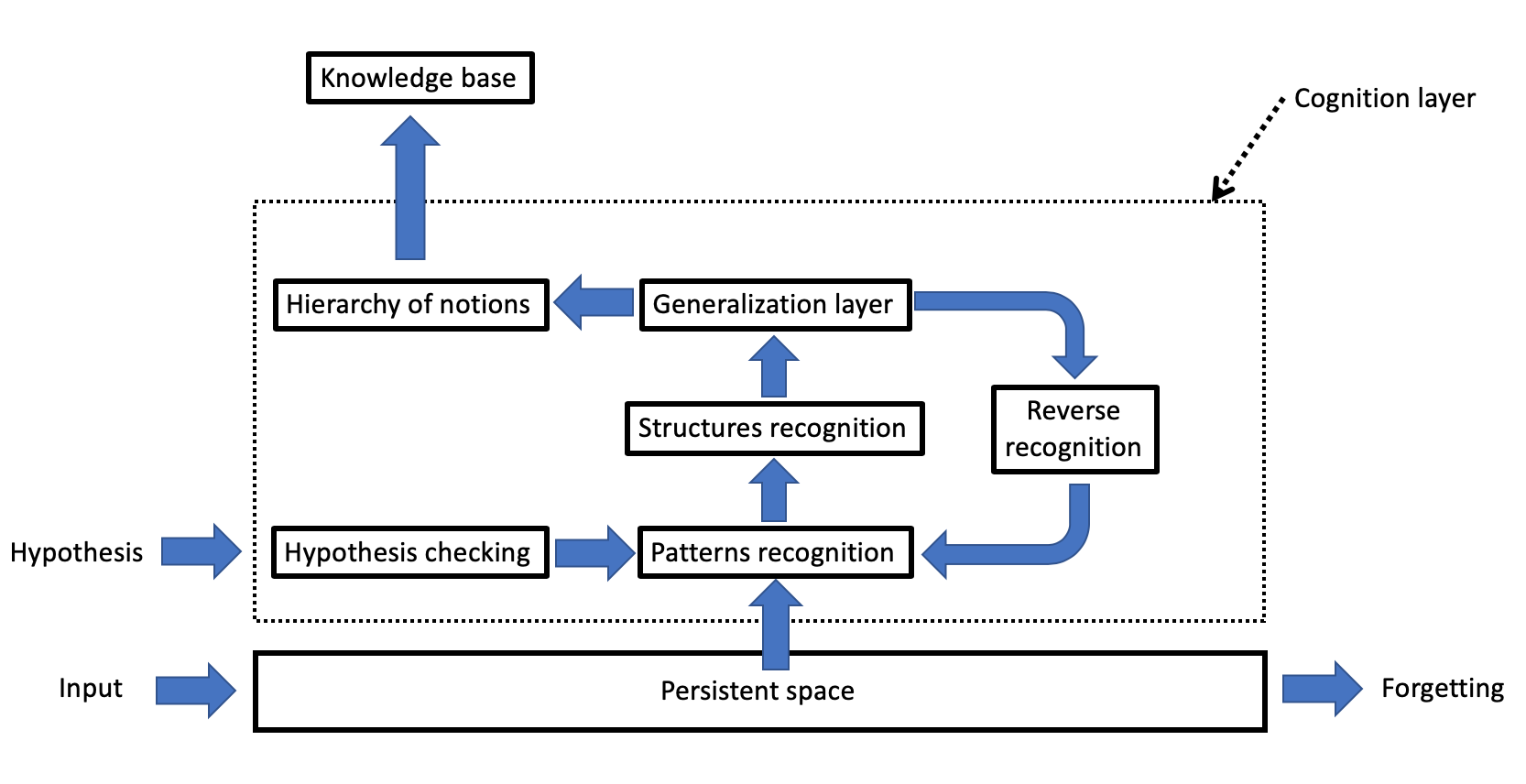}
\caption{Cognition system architecture.}
\label{fig:fig-00006}
\end{figure}

Moreover, the probability (or another factor) could estimate the correctness of the hypothesis. And such probability could be used for correction or rejection of the hypotheses. The ability to forecast future states of input stream employing hypotheses generation could be a steady basis for knowledge synthesis and the elaboration of the system's behavior. If the system can synthesize some hypothesis and this synthesized pattern detected in the input stream, then it means that the system discovered a feature of the reality in the form of synthesized knowledge. Moreover, if the system expects some future state of the input stream, then it is possible to prepare some resources for reaction on this predicted state of reality.

\section{WHAT IS THE RESULT OF COGNITIVE COMPUTING?}

The cognitive computing paradigm has to implement the mechanism of synthesizing conclusions based on available data. Generally speaking, the whole workflow of cognitive computing includes several steps, but every action has its result. Finally, as a result, the workflow should implement the checking, probability determination, and relevancy elaboration of the synthesized hypotheses. Moreover, the assumption represents a generated statement is incorporating the available data. Such statements are the new data that, potentially, can predict or forecast the future states of the input data stream. As a result, the checked and confirmed hypotheses could play the role of real knowledge that needs to place into the knowledge base.

\subsection{Patterns recognition}

If anyone tries to imagine the complete chaos, then it is possible to explain the entire mess as the impossibility of encountering a repeatable pattern in a data stream. However, the bytes of data can contain only 256 various values. Generally speaking, it means that probability to discover the repeatable byte's value is very high for any data stream. The Universe has such a structure that humankind is capable of detecting the knowledge about reality based on repeatable patterns of data. It means that if anybody considers a raw binary stream of data is representing a state of existence, then the probability of encountering the repeatable patterns of information is very high. Generally speaking, the cognitive computing paradigm could implement the analysis of raw data streams through the discovery of repeatable patterns of data. Such a recognition function needs to be the first step of the whole paradigm. Because the possibility to recognize the repeatable patterns of data is the basis of reality's investigation and elaboration of Nature's laws by humankind. Moreover, the distinction of repeatable patterns is the crucial technique for the extraction and building of the understanding of abstract keywords or notions in the input data stream.

\subsection{Structures recognition}

Generally speaking, the repetition of patterns allows detecting it in the input data stream. Moreover, patterns do not exist alone. Usually, pieces of data join into the structures, and these structures identified in the information data streams. Finally, structures of patterns provide the opportunity to discover some notions in data streams that represent the essence of processes that are taking place into reality. Also, any structure is a set of ordered patterns (keywords) located on some positions in the sequence. As a result, a graph can represent a structure's organization.

Moreover, comparing the structures with each other is the possible way to detect the identity or similarity of structures. The similarity or identity of patterns and the ordering of patterns in the sequence is such foundation. Generally speaking, the opportunity to discover the various environments for patterns existence defines the importance of structure. The structure reveals the relations among the different patterns. Also, it detects the possible ties of notions in reality. Finally, the knowledge of relationships among the patterns provides a way to generalize the concepts and to synthesize the new statements.

\subsection{Notions generalization}

One of the essential nature of humankind's culture of thinking is the ability to generalize notions. The generalization of concepts is the fundamental basis of abstract thought and the deducing of generalized conclusions. The cognitive computing paradigm is capable of mimicking the human's ability to generalize the notions. And the similarity of structures could be the basis for such generalization. For example, if anyone imagines two different specimens of structure that differ by one pattern in some position, then it could mean the synonymy of patterns. To be more precise, it implies that two particular patterns represent the synonymy with some abstract notion. Generally speaking, the structure can be generalized through exchanging the discovered position by abstract keyword. Any specific keyword can substitute this abstract keyword or location is detected previously. As a result, two particular specimens of structures are capable of being "generalized" to an abstract composition that contains the "notion(s)" instead of specific keywords (patterns).

Moreover, the generalized position will be associated with the vector of discovered particular keywords. Such generalization of specific patterns could be continued until the recognition of all patterns in the structure and converting the structure's positions into the abstract keywords are associated with the vectors of particular keywords. Moreover, it is possible to say that the generalization step provides the basis for discovering the abstract notions and the way to recognize the synonymy or similarity of patterns. Also, generalized structures allow revealing the abstract relations of patterns and general rules of structure synthesis.

\subsection{Hierarchy of notions creation}

It is essential to point out that the generalized structure is capable of associating several particular structures. Generally speaking, it means that such a generalized structure could look like a parent node for these specific leaf structures. Moreover, these parent and leaf structures build the hierarchy of nodes that store the knowledge of a way to generalize any particular structure into a generalized one. And if anyone continues the process of structure generalization, then it is possible to build the multi-level hierarchy of abstract notions. As a result, the generalized formulas could be a basis for statements synthesis. Generally speaking, it is possible to state that notions hierarchy creates the structures' (or statements') relations. Finally, this net of relations builds a "representation of reality" in the form of generalized notions. The generalization of concepts introduces the classification of groups the particular keywords into the generalized classes.

Moreover, discovering the generalized classes of notions is the first step towards a paradigm of "artificial analytical thinking". This approach implies that the system will be converted into the "thinking machine" by deducing the input data stream's particular notions into the abstract classes. Generally speaking, it implies the opportunity to classify data in the input binary stream by the system itself in an automatic mode based on the created hierarchy of notions. The other important responsibility of hierarchy is the synthesis of new statements and new knowledge. Generally speaking, the notions hierarchy is based on discovering the similarity among pattern's structures and, as a result, the generalization of notions. Finally, the side effect of such generalization is converting the particular structures into the abstract formulas. These formulas can build new statements employing substitution of the generalized positions by any key/pattern from the same or another vector of keywords. However, what could be a possible direction for statement synthesis? And it is not a rhetorical question because the notions hierarchy is capable of including thousands of abstract formulas. First of all, it is possible to point out that the synthesis of a statement could use an existing informational context. It means that the input stream or user's requests could define a set of relevant keywords or patterns. Generally speaking, the proper set of keywords can select a vector of particular keywords to use some abstract formulas from the notions hierarchy. Finally, these selected formulas could be used for the synthesis of the new hypothesis by relevancy to the current informational context. Also, the synonymy of patterns/keywords can play a crucial role during the statement synthesis process.

\subsection{Multi-dimensional relations creation}

If anyone considers an aggregation of particular leaf structures that are associated with an abstract parent, then it is possible to see the similarity between this case and the relational model. Generally speaking, it is possible to consider the set of particular structures like records of a column in a table. Moreover, the process of notions generalization has a side effect of naturally building the relational model of available data. In the case of the cognitive computing paradigm, the system itself is synthesizing the relational model, and it will evolve with the changing nature of the input data stream. Also, the data structuring and ordering is the vital responsibility of any data storage system.

The unstructured data model is the most flexible scheme of data representation. However, it is a nightmare for any system that implements SQL-like requests. The cognitive computing paradigm provides a way to store the data without any initial structuring. However, this model can synthesize the relational model of available data during the data analysis and generalization. Generally speaking, the relational model is the side-effect of the creation of a hierarchy of generalized notions. First of all, most abstract concepts will be at the top of the hierarchy. As a result, keywords in a search request can select some subset of particular notions or structures that are associated with the requested abstract concept.

Moreover, if anyone creates the request is containing the set of abstract keywords, then the execution of such request retrieves the union or join of subsets of particular structures. These structures represent the child nodes for the requested abstract notions. Also, it is possible to use the hybrid way of search where it is possible to generalize the initial keywords in the request into the abstract notions. Finally, these abstract notions could be the basis for a more wide search request. And it is possible to deduce the result of such search into the particular structures representing the specific notions. The potential advantage of such type of request could be the using of synonymy of notions that provides the way to retrieve more relevant and useful data as a result of requests.

\subsection{Data relevancy elaboration}

The hierarchy of notions creates the foundation for analytical thinking. However, this hierarchy is only the space that can classify and order all available and existing notions. But it needs to introduce relevancy for converting the hierarchy of notions into the more powerful concept. Generally speaking, the relevancy defines the probability of encountering a notion into the future states of the data stream or user requests. From one point of view, any data system exists intending to satisfy the users' demands. It means that any keyword or pattern has a high relevancy rate if a user needs any information related to this notion. However, from another point of view, if the input data stream contains a pattern with some frequency, then such a keyword is relevant because nobody can ignore reality. As a result, it is possible to say that relevancy defines the direction of the system's evolution.

If the system needs to work without taking into account the relevancy, then all keywords or patterns are equal, and it implies the necessity to evolve in all possible directions. However, the system's resources are always limited, and any user is interested in data that are relevant to current tasks. It means that the user stores data and creates requests that are relevant to the current problems and needs. Moreover, if a system provides the services to end-users, then, anyway, the data stream will include some patterns with higher frequency comparing with other ones. Generally speaking, the probability of detecting a pattern in the data stream defines the relevancy of this keyword.

Moreover, the notion's or keyword's relevancy could play a critical role in the synthesis of new statements or pre-views preparing. But the relevancy is capable of being a more crucial point in the system because of the ability to elaborate on the system's behavior strategy. If anyone imagines that a machine is responsible for self-creation of analysis and synthesis processes, then it emerges the question - what should be a foundation for generation any simple act of analysis? Practically, it is possible to initiate the analysis or synthesis process for all detected patterns or keywords. But such an approach cannot be considered as a feasible one by quickly exhausting all available resources without achieving any practically useful results. Generally speaking, relevancy is the practical basis for the system's resources distribution and to work out some valuable results in the form of synthesized statements or conclusions during the limited time.

\subsection{Hypothesis synthesis and elaboration}

The most critical possible responsibility of cognitive computing could be the classification of data and recognition of input data stream's states. Moreover, the capability to predict the future states of the input data stream could be a viable feature. The synthesis of new statements creates the foundation for the foreseeing the possible data states and elaboration of knowledge about new data classes. Generally speaking, the formulation of a new understanding means to detect the laws of reality through discovering the statements' formulas in the environment of the available data stream. As a result, a piece of new knowledge can be identified by employing synthesis and checking the hypothesis. Any synthesized statement or formula that was never encountered before in the input data stream looks like a hypothesis.

It is possible to distinguish two abstract points of knowledge about the input data stream. The first point is the complete absence of knowledge about the input stream. Oppositely, the second point is the knowledge of all dependencies and relations in the input data stream. As a result, the complete absence of knowledge about the input stream implies that it is always possible to discover any new patterns never detected before. The point of holding full experience means the capability to classify all detected patterns in the input data stream. Generally speaking, the hypotheses synthesis should take place under the necessity to deduce and to elaborate the possible classes of patterns in the input data stream and, finally, to build the knowledge of notions' dependencies in the informational reality.

Moreover, hypotheses synthesis is a technique of pro-active thinking. It means that hypotheses synthesis is capable of forecasting and of synthesizing the statements or formulas that cannot be deduced directly from the available data set. For example, the detection of variation of patterns in a similar structure could look like not a complicated task. However, the task of generalization of the several patterns in a structure to build the abstract notions hierarchy could be entirely not a trivial task that cannot be solved by merely matching the available patterns of data. Namely, the hypotheses synthesis could be the technique of resolving the problem of abstract notions hierarchy creation and evolution.

Generally speaking, a hypothesis looks like any statement that it needs to check for correctness. It means that in the cognitive computing paradigm, a hypothesis could be synthesized based on the conversion of generalized statement's formula into the particular statement employing substitution the generalized notions by specific keywords from the vector(s) or bag(s) of keywords. Moreover, it is possible to synthesize the hypotheses to interpolate the generalized structures into abstract notions. Also, a user request can be the basis for building the hypotheses that are relevant to the user's request.

Also, it is possible to build a hypothesis to create a new structure. The similarity of structures could be a basis for such synthesis. It means that a new structure can combine the elements of similar structures by employing the variation of items in it. In any case, the goal of hypotheses synthesis is the generation of new generalized knowledge about the reality in the environment of the pro-active "thinking" to forecast the future states of the input data stream and to elaborate a strategy of the system's behavior.

\subsection{Hypothesis checking, correction and rejection}

The hypotheses synthesis is the first step towards a piece of new knowledge, but it needs to check any inference before transforming it into the experience. It means that it needs to analyze the input data stream to discover a pattern that is similar or identical to the synthesized hypothesis. Generally speaking, if it is possible to detect any structure or pattern in the input data stream that mirrors the integrated statement, then the assumption is correct, and it is a piece of new knowledge about the reality. Finally, it implies the hypothesis can be transformed and stored into the hierarchy of notions as real knowledge.

However, if the hypothesis and the pattern of the input data stream are not identical, then it needs to estimate the deviation of the assumption and, maybe, to fulfill the hypothesis correction. Generally speaking, it needs to employ some method that can calculate the "distance" between the hypothesis and the detected pattern or structure. The calculated "distance" can reveal the deviation strength, but it cannot define the necessary correction of the synthesized hypothesis. It needs to take into account the several cases of deviation for the same assumption that could determine the required revision of the hypothesis. The synthesized hypothesis could include several new items (patterns) injected during the synthesis process. And the detected deviation can identify the elements that create the hypothesis's bias. Finally, these items could be the reason for the inadequacy of the synthesized assumption, and they should be corrected wholly or partially.

Generally speaking, the correction implies the creation of a new, corrected hypothesis that should have a shorter "distance" with the known patterns in the input data stream. As a result, new detected patterns in the input data stream could show that: (1) hypothesis becomes closer to the real knowledge; (2) assumption has more inadequacy after the correction; (3) hypothesis fluctuates near some deviation point. It is possible to correct the hypothesis through the exclusion of some items. Also, the "fluctuating" hypothesis needs to exclude as not to have enough potential to achieve the state of real knowledge. If the input data stream does not contain any similar patterns for a hypothesis, then this situation could be treated in two possible ways. Generally speaking, the assumption could be wrong, or there is no data in the input data stream to confirm or check the hypothesis. Finally, it is possible to use some timeout for waiting for the data that can prove the correctness of the hypothesis. Oppositely, it could be more productive and efficient to reject the hypothesis and to synthesized it again in the future.

\subsection{Behavior strategy elaboration}

Usually, any system receives the input data stream that is a snapshot of reality - however, the persistent space stores this binary stream of data without any initial interaction with the system. Generally speaking, the system begins to analyze the data of being stored in a persistent space. However, the system needs to interact with end-users while it is investigating the input data stream in the background. Because, finally, the goal of any information system is to process the end-users' requests and to provide data as the result of requests fulfillment. Finally, it is possible to say that system has several channels of interaction with the end-users and ever-changing reality.

The system could be capable of distinguishing the patterns or keywords in end-users' requests and identifying the commands and the keywords, and it can use it during the search and requests execution. However, even a single-user system can receive the volume of requests that could require the computing power out of an available system's resources. Generally speaking, it needs to consider the elaboration of some strategy that could provide the opportunity to fulfill the maximal possible users' requests in the environment of limited system resources. It means that system has to select such a subset of users' requests that can be generalized and to be executed in a generalized form with the goal to provide the relevant data expected by users. The hierarchy of abstract notions provides a way to generalize the particular keywords in the users' requests to the abstract ones.

It is possible to point out that the generalization of keywords is an essential step because it decreases the number of particular requests by means of merging several particular requests into the generalized one. Finally, it is possible to split the whole set of requests on the several generalized subsets and to prioritize it based on the relevancy of generalized keywords. It is possible to assign the highest priority to a generalized request that is able to process the biggest number of particular requests. However, the result of a generalized request needs to be processed additionally for every specific request with the goal to increase the relevancy of data.

Generally speaking, such a technique can decrease the number of requests by executing the initial pre-view preparation. Also, it can increase the parallelism of transformation the result of generalized request into the relevant data set for particular requests. Additionally, the elaboration of behavior strategy could imply the dialog between the system and the end-user to make a request more specialized. Such dialog could be used by the system to make very abstract user's requests more particular or to interact with the user during the request execution.

\subsection{Forecast the future state of input data stream}

The forecast of input data stream's future states could be the crucial feature of cognitive computing. But what could be implied by the foreseeing of future states? Generally speaking, it is possible to suggest a synthesis of the input data stream's future states as the forecast activity. However, it could not be so valuable and resource-consuming activity. Moreover, it is possible to imagine some classes of users' requests that inquire about a potential value of some measurement or estimation in the future.

Usually, end-user issues the claims for existing data that mirror our knowledge about the past or the present. But similar requests could be questioned for the future states of data also. For example, it is possible to ask about "how many people will play football in the future?". This request looks like a regular request, but it is expected some forecast as a result of the request execution. Generally speaking, any user's request is a set of keywords that restrict the searching space. It means that the request's keywords are capable of defining a leaf set of structures relevant to the case of a claim.

Moreover, a multi-dimensional relation model of abstract notions can order the leaf structures based on the timestamp or any other factor. As a result, it implies the opportunity to employ such an ordered sequence for the prediction of possible future states. If timestamps order the leaf structures, then it means that such an ordered sequence reveals the evolution of leaf structures. Generally speaking, it is possible to elaborate a vision of the future state based on the timestamp-ordered chain. Finally, this way of considering the timestamp-ordered sequence provides the opportunity to synthesize the hypothesis.

Generally speaking, any set of records can be classified. It implies that it is possible to split a sequence of particular structures or records into several groups or classes. As a result, every leaf record in the chain can be a member of some known class. And it is possible to estimate the number of particular structures or members in every class. Finally, the number of members in one category can be a quantitative feature of the class. But the distribution of classes in the sequence is a qualitative feature of the sequence's evolution. To conclude, if it is possible to discover some vision of a sequence's evolution by quantitative and/or qualitative analysis, then it is possible to elaborate a view of the sequence continuation and to synthesize a potential future classes distribution in the sequence.

\section{WHAT ARE THE ADVANTAGES AND DISADVANTAGES OF COGNITIVE COMPUTING?}

\subsection{Potential advantages}

\textbf{Capability to recognize the repeatable patterns.} The humankind’s ability to cognize the Universe is the capability to recognize the repeatable patterns of the reality. Generally speaking, computer system is capable to mimic the technique of knowledge extraction from the raw data through employing the repeatable patterns recognition. The technique of repeatable patterns detection is simple and it does not require the extensive computational power. As a result, it means that this technique is not power-hungry. However, the mechanism of repeatable patterns recognition could create the basis for sophisticated approaches of data analysis and knowledge extraction without the using of algorithms are created by humans.

\textbf{Data self-assembling.} Generally speaking, it is possible to consider the patterns recognition in a raw data like the keywords distinguishing in the data stream. It means that this technique can be the basis for conversion of raw data stream in the KV store representation. Moreover, it is possible to employ the possibility to detect the identity or similarity of keywords for building the relations among the data items. As a result, these relations can create hierarchical or relational data model. It is crucial to point out that computer system will be capable to detect and create the relations without using any specialized algorithms created by humans. Generally speaking, the system can recognize the data organization and to transform the unstructured data into the relational model by itself without any special hints. Moreover, system will be capable to rebuild the data model dynamically when a new data will be stored into the persistent space.

\textbf{Capability to generalize notions.} The nature of human culture is the ability to create the abstract or generalized ideas. Namely abstract concepts make the foundation of human culture to think. Generally speaking, cognitive computing paradigm can introduce a mechanism of notions generalization. The basis of such technique is the detection of data structures similarity that can discover the identity or similarity of keywords’ senses. As a result, synonymy can build hierarchy of keywords from particular to abstract notions. Finally, the generalized notions can play crucial role in data analysis and knowledge synthesis.

\textbf{Self-assembling of multi-dimensional data relations.} The reality includes complex and multi-dimensional relations among the subsystems. It is possible to deduce this conclusion from the point of multi-dimensional and multi-relational nature of items in the data stream. Generally speaking, the cognitive computing paradigm represents the tool is capable to discover the multi-dimensional data bonds and to transform it into multi-dimensional relational model. Moreover, the system can rebuild the synthesized data model dynamically if the input data stream introduces new classes of data. It is important to point out that analyzing the multi-dimensional relations amongst the data is very complex problem for the human brain. Finally, it means that cognitive computing paradigm can be crucial and very profound tool for analyzing complex and multi-dimensional data relations.

\textbf{Capability to detect the relevancy of data.} Generally speaking, the cognitive computing paradigm has goal to cognize a new knowledge analyzing the input data stream. However, the data stream could contain as relevant as useless data from the end-use point of view. As a result, it implies that the system would waste the computational power without the determination of data relevancy. Finally, the system could lose the energy, free space, and to introduce the unexpected lag of user’s requests execution. Moreover, theoretically, the whole system could be busy by processing not relevant data that is able to result in declining to fulfill the user’s requests or to execute the user’s requests in the background.

The capability to generalize notions and to make the self-assembling of data is the foundation of cognitive computing paradigm. Generally speaking, it implies the ability of system to recognize the repeatable patterns in the input data stream and to estimate the relevancy of recognized data structures on the basis of frequency to detect this or similar data structure in the input data stream. As a result, the knowledge of frequency to encounter a data structure in the past states of data stream provides the basis to estimate the relevancy of some data in the future states of input data stream. Moreover, the knowledge of relevancy a data structure or pattern creates the basis to correct the user’s requests into more relevant state to generalize or concretize the initial request.

\textbf{Capability to synthesize and to check hypotheses.} The analysis of input data stream is able to fulfill the data self-assembling and deducing the generalized notions. However, the hypotheses synthesis and checking could be the basis for new knowledge elaboration. Generally speaking, a hypothesis is an initial statement that could be true or false one. The hypothesis check results in rejection of the statement or adoption it as a new knowledge.

Cognitive computing paradigm is capable to evolve and to synthesize the knowledge on the basis of hypotheses generation and checking. Generally speaking, combination of generalized data structure with particular data patterns is the way of synthesizing a hypothesis. The high relevancy is the possible criterion for selection of generalized data structure. Additionally, it is possible to consider a similarity of data structure with unrecognized data pattern as the complimentary criterion. Finally, the examination of knowledge base or the process of parsing the input data stream is able to discover the true or false state of the hypothesis.

\textbf{Capability to synthesize a new data.} The necessity to rebuild a scheme of any relational database in the case of adding a new type or class of data is the crucial drawback of the relational model. Moreover, such database refactoring could be very complex or practically impossible if database contains significant amount of data. Anyway, the refactoring of database scheme is very time-consuming process.

The advantage of cognitive computing paradigm is the capability to build and to refactor the scheme of data representation and ordering dynamically. Generally speaking, the system is able to elaborate and to rebuild the scheme of data representation by itself without involving the human participation or any human activity. Moreover, such synthesized scheme and the generalized notions create the way of data synthesis by means of combination the generalized structures with particular notions by using the synonymy of keywords.

\textbf{Capability to forecast the possible future states of input data stream.} The goal of any analysis is the elaboration of a consistent pattern that could provide the basis for prediction of future states the input data stream. Cognitive computing paradigm is the powerful tool for solving this task. The ability to detect the relevance of data provides the opportunity to discover data structures and data patterns that future state of data stream could include. Moreover, notions generalization is the basis for hypotheses synthesis that could represent potential data patterns in the future states of input data stream. However, the correctness of hypotheses will rely on knowledge base’s completeness and relevance of available data in the system.

\subsection{Potential disadvantages}

\textbf{Complexity of performance estimation.} Potentially, cognitive computing paradigm provides profound opportunities for massively parallel data processing. Because, it is easy to split the input data stream amongst the set of analyzing threads in the scope of paradigm. Moreover, the recognized patterns could originate the independent threads of analysis and synthesis of data. However, the opportunity to process data in massively parallel and decentralized manner could be considered like the potential issue of threads management. Generally speaking, end-user could consider the absence of centralized management like the reason of performance degradation and not efficient employing of system’s resources. Moreover, decentralized model of data processing activity in cognitive computing paradigm makes the performance estimation by probabilistic value. It means that a request’s performance estimation could vary dramatically because internal system’s activity could significantly affect the measurement act.

\textbf{Complexity to estimate and to manage the energy consumption.} It is well known fact that every computing activity consumes the energy. If anybody considers a system under centralized management, then it is a deterministic system that spends energy only when end-user requests the calculation. Generally speaking, the user is the source of activity in the system and the reason of energy consumption.

However, cognitive computing paradigm is the decentralized technique where the system can initiate the processes of data analysis and synthesis. Moreover, it implies that the volume of data on the system could define the amount of internal calculation activity in the system. Generally speaking, it means that, potentially, the system could consume a significant amount of energy even without any user requests. Finally, it could be not easy to estimate the energy consumption in the environment of cognitive computing paradigm.

\textbf{Unintentional "hiding" data by system.} The system builds a data structure by itself. As a result, it implies potential inability to recognize some patterns or data structures because of incomplete knowledge base or malfunctioning. Generally speaking, it means that the system could "hide" some data unintentionally. If anybody considers a traditional relational database then such knowledge base is able to contain and to provide access to data that satisfies to requirements of database’s scheme. As a result, the scheme guarantees that stored data will be accessed by any request in the case of successful data aggregation into the database.

However, database scheme makes the database by completely not flexible entity. It means that necessity to integrate any new class of data implies the development activity by skilled personnel. Generally speaking, such database modification can be time-consuming operation or even not feasible option. Cognitive computing paradigm suggest the way to build the database's scheme by system itself. Moreover, system is rebuilding the scheme if any new class of data is detected by the system. However, the potential side effect of such flexibility could be the inability to recognize some class of data by virtue of incomplete knowledge base.

\textbf{Incorrect estimation of data relevancy.} The cognitive computing paradigm is based on data relevancy estimation. Finally, it means that incorrect relevancy estimation leads to incorrect search results or hypotheses synthesis. But the relevancy estimation is always probabilistic process and it implies the significant probability of erroneous estimations. Generally speaking, it could imply that a particular user request can be estimated like not relevant and it will be executed by the system in the background. As a result, the end user may estimate the system performance as poor and inadequate to the importance of the request. Moreover, it implies that system will schedule not enough resources or no resources at all to process not relevant data request. Finally, it could result in ignoring such data and not integrating these data into the knowledge base. As a result, even if system has the necessary data then, anyway, a result of user request could not include it. Generally speaking, it means that erroneous estimation of data relevancy could result in incorrect hypotheses that cannot predict the future states of input data stream. However, decentralized model of cognitive computing paradigm is able to provide the basis for competing the DPU cores for data processing and analysis. As a result, it means thousands of independent and competing DPU cores is capable to resolve the potential problem of incorrect data relevancy estimation.

\textbf{Potential conflict of requests' priority.} If system estimates the priority of requests then the presence of many users and/or many requests could create the conflict of requests' priority. It means that system decision could conflict with the end-users expectations. However, if the system has enough relevant data and user requests access the relevant data then the probability of such conflicts could be very low. Generally speaking, potentially, really fresh data could be not treated as relevant yet and it can create the basis for such conflicts. Finally, many users in the system are capable to create the environment where could be not easy to estimate the request priority properly. However, the cognitive computing paradigm is the data-centric approach that allocates relevant amount of resources for a data volume is participating in the request execution.

\textbf{Decentralized nature of execution.} Generally speaking, decentralized execution of requests could potentially result in infinite cycles, deadlocks, race conditions and likewise side effects. However, it is not so definite conclusion that decentralized model of execution is a basis for various synchronization issues for the case of shared data. For example, if anybody considers a DPU that owns a portion of data then one data owner is capable to guarantee the correctness of data access and modification in multi-core or multi-threaded environment. Such guarantee can be implemented through requests queue that is able to serialize requests from many actors. From another point of view, if anybody imagines a DPU matrix is receiving a request that should be processed in decentralized manner then, potentially, such request could travel through the DPUs set infinitely. Generally speaking, it needs to declare a very clear protocol of DPUs interactions that can exclude the infinite cycles or deadlocks. Finally, cognitive computing paradigm requires dedicated protocol of integration and coordination of DPUs activities and interactions.






\end{document}